\documentclass[fleqn,usenatbib]{mnras}

\usepackage{newtxtext,newtxmath}

\usepackage[T1]{fontenc}

\DeclareRobustCommand{\VAN}[3]{#2}
\let\VANthebibliography\thebibliography
\def\thebibliography{\DeclareRobustCommand{\VAN}[3]{##3}\VANthebibliography}

\usepackage{graphicx}	
\usepackage{amsmath}	

\title[A $z=0.7092$ OH megamaser in the MIGHTEE survey]{The discovery of a $z=0.7092$ OH megamaser with the MIGHTEE survey}

\author[M. J. Jarvis et al.]{Matt J. Jarvis$^{1,2}$\thanks{E-mail: matt.jarvis@physics.ox.ac.uk},
Ian Heywood$^{1,3,4}$,
Sophie M. Jewell$^{1,5}$,
Roger P. Deane$^{6,7}$, 
H.-R.~Kl\"ockner$^{8}$,
\and 
Anastasia A. Ponomareva$^{1}$,
Natasha Maddox$^{9}$, 
Andrew J. Baker$^{10,2}$,
Alessandro Bianchetti$^{11,12}$,
\and
Kelley M.\ Hess$^{13}$,
Hayley Roberts$^{14,15}$,
Giulia Rodighiero$^{11,12}$,
Ilaria Ruffa$^{16}$,
Francesco Sinigaglia$^{11,12,17}$,
\and
R. G.\,Varadaraj$^{1}$,
I.H. Whittam$^{1}$,
Elizabeth A.\ K.\ Adams$^{18,19}$
Maarten Baes$^{20}$,
\and
Eric J.\,Murphy$^{21}$,
Hengxing Pan$^{1}$,
Mattia Vaccari$^{22,23,24}$
\\
$^{1}$Astrophysics, Department of Physics, University of Oxford, Keble Road, Oxford OX1 3RH, UK\\
$^{2}$Department of Physics and Astronomy, University of the Western Cape, Robert Sobukwe Road, 7535 Bellville, Cape Town, South Africa\\
$^{3}$Department of Physics and Electronics, Rhodes University, PO Box 94, Makhanda, 6140, South Africa \\
$^{4}$South African Radio Astronomy Observatory, 2 Fir Street, Black River Park, Observatory, Cape Town 7925, South Africa\\
$^{5}$Institute for Astronomy, Royal Observatory Edinburgh, Blackford Hill, Edinburgh, EH9 3HJ, UK\\
$^{6}$Wits Centre for Astrophysics, School of Physics, University of the Witwatersrand, 1 Jan Smuts Avenue, Johannesburg, 2000, South Africa \\
$^{7}$Department of Physics, University of Pretoria, Private Bag X20, Pretoria 0028, South Africa\\
$^{8}$Max-Planck Institut f\"{u}r Radioastronomie, Auf dem H\"{u}gel 69, 53121 Bonn, Germany\\
$^{9}$School of Physics, H.H. Wills Physics Laboratory, Tyndall Avenue, University of Bristol, Bristol, BS8 1TL, UK\\
$^{10}$Department of Physics and Astronomy, Rutgers, The State University of New Jersey, 136 Frelinghuysen Road, Piscataway, NJ 08854-8019, USA\\
$^{11}$Department of Physics and Astronomy, Università degli Studi di Padova, Vicolo dell’Osservatorio 3, I-35122, Padova, Italy\\
$^{12}$INAF - Osservatorio Astronomico di Padova, Vicolo dell’Osservatorio 5, I-35122, Padova, Italy\\
$^{13}$Department of Space, Earth and Environment, Chalmers University of Technology, Onsala Space Observatory, 43992 Onsala, Sweden\\
$^{14}$ School of Physics and Astronomy, University of Minnesota, 116 Church St. SE, Minneapolis, MN 55455, USA\\
$^{15}$Minnesota Institute for Astrophysics, University of Minnesota, 116 Church St. SE, Minneapolis, MN 55455, USA\\
$^{16}$Cardiff Hub for Astrophysics Research \&\ Technology, School of Physics \&\ Astronomy, Cardiff University, Queens Buildings, The Parade, Cardiff, CF24 3AA, UK\\
$^{17}$Départment d'Astronomie, Université de Genève, Chemin Pegasi 51, 1290, Versoix, Switzerland\\
$^{18}$ASTRON, the Netherlands Institute for Radio Astronomy, Oude Hoogeveesedijk 4,7991 PD Dwingeloo, The Netherlands\\
$^{19}$Kapteyn Astronomical Institute, University of Groningen, PO Box 800, 9700 AV Groningen, The Netherlands\\
$^{20}$Sterrenkundig Observatorium, Universiteit Gent, Krijgslaan 281 S9, B-9000 Gent, Belgium\\
$^{21}$National Radio Astronomy Observatory, 520 Edgemont Road, Charlottesville, VA 22903, USA\\
$^{22}$ Inter-University Institute for Data Intensive Astronomy, Department of Astronomy, University of Cape Town, 7701 Rondebosch, Cape Town, South Africa\\
$^{23}$Inter-University Institute for Data Intensive Astronomy, Department of Physics and Astronomy, University of the Western Cape, 7535 Bellville, Cape Town, South Africa\\
$^{24}$ INAF - Istituto di Radioastronomia, via Gobetti 101, 40129 Bologna, Italy
}

\date{Accepted XXX. Received YYY; in original form ZZZ}

\pubyear{2015}

\begin{document}
\label{firstpage}
\pagerange{\pageref{firstpage}--\pageref{lastpage}}
\maketitle

\begin{abstract}
 We present the discovery of the most distant OH megamaser to be observed in the main lines, using data from the MeerKAT International Giga-Hertz Tiered Extragalactic Exploration (MIGHTEE) survey. At a newly measured redshift of $z = 0.7092$, the system has strong emission in both the 1665\,MHz ($L \approx 2500$~L$_{\odot}$) and 1667\,MHz ($L \approx 4.5\times10^4$~L$_{\odot}$) transitions, with both narrow and broad components. We interpret the broad line as a high-velocity-dispersion component of the 1667\,MHz transition, with velocity $v \sim 330$\,km~s$^{-1}$ with respect to the systemic velocity. The host galaxy has a stellar mass of $M_{\star} = 2.95 \times 10^{10}$\,M$_{\odot}$ and a star-formation rate of SFR = 371\,M$_{\odot}$\,yr$^{-1}$, placing it $\sim 1.5$\,dex above the main sequence for star-forming galaxies at this redshift, and can be classified as an ultra-luminous infrared galaxy. Alongside the optical imaging data, which exhibits evidence for a tidal tail, this suggests that the OH megamaser arises from a system that is currently undergoing a merger, which is stimulating star formation and providing the necessary conditions for pumping the OH molecule to saturation. The OHM is likely to be lensed, with a magnification factor of $\sim 2.5$, and perhaps more if the maser emitting region is compact and suitably offset relative to the centroid of its host galaxy's optical light. 
This discovery demonstrates that spectral line mapping with the new generation of radio interferometers may provide important information on the cosmic merger history of galaxies.

\end{abstract}

\begin{keywords}
masers -- ISM:molecules -- galaxies:ISM -- galaxies:starburst
\end{keywords}



\section{Introduction}

Hydroxyl masers were discovered over five decades ago, with the majority of early detections coming from compact H{\sc ii} regions in our own Galaxy \citep{Weaver1965}. \cite{Perkins1966} were the first to interpret these lines as maser emission, which provided an explanation for the high brightness temperature, polarisation properties and the line ratios. The following year,
\cite{WilsonBarrett1968} discovered OH emission from four infrared stars, although many more did not exhibit detectable OH emission. \cite{Heiles1968} also detected OH emission from interstellar dust clouds, suggesting that OH emission arose from regions with a large preponderance of infrared emission. As observations of the sky at radio wavelengths became more widespread, OH emission was discovered in external galaxies \citep[e.g.][]{BaanWoodHaschick1982}.  These masers tended to be extremely luminous, and were referred to as megamasers due to them being over a million times more luminous than typical Galactic interstellar OH maser sources \citep[see][for a review]{Lo2005}. 

The OH molecule has four hyperfine transitions due to the coupling of the spin of the unpaired electron with the nuclear spin of the hydrogen atom. These transitions occur at 1612, 1665, 1667, and 1720\,MHz, with line ratios of 1:5:9:1 in local thermodynamic equilibrium. The conditions necessary for maser emission include a source of energy that ensures that there are more molecules in the upper energy level than in the lower, in order to produce the stimulated emission. Indeed, it has become clear over the past few decades that OH megamasers (OHMs) are closely associated with galaxies with significant infrared emission, exhibiting a tight correlation between the far-infrared luminosity and the OHM luminosity \citep[e.g.][]{Baan2008,Wang2023}. Theoretical models have shown that OHM emission is efficiently produced at relatively high temperatures ($\sim 80-140$~K), with a minimum temperature of $\sim 45$\,K
 needed for inversion \citep{LockettElitzur2008}. However the dust around star-forming regions in external galaxies tends to be much cooler than this ($30-50$~K) \citep[e.g.][]{Hwang2010,Smith2013,Smith2014}. Thus, to produce maser emission in these cooler environments means that the OH gas is likely co-located with the heating source. \cite{Willett2011a, Willett2011b} investigate the dependence of the OHM luminosity on the mid-infrared emission using data from the Spitzer Infrared Spectrograph to argue that a large smoothly distributed dust reservoir with temperatures from $\sim 50-100$\,K and high opacity ($\tau \sim 100 - 400$) is required for OHM emission.

However, there are key differences between using an infrared galaxy survey as a parent sample to look for masers \citep[e.g.][]{Baan1992,Staveley-Smith1992,Norris1989}, and a purely spectral line survey that can detect OHMs irrespective of the host galaxy properties \citep[see e.g.][]{Townsend2001}. It is only the latter that allow us to understand the full-range of conditions sufficient to produce OHM emission.

To carry out such surveys, one needs large spectral bandwidths at radio wavelengths, where the OH emission is detected "blindly", with no pre-selection of samples such as (Ultra)Luminous Infrared Galaxies (U)LIRGS and/or mergers \citep[e.g.][]{DarlingGio2006}. Many of the most successful untargeted surveys for OHMs are nominally focussed on detecting and understanding neutral atomic hydrogen via the 21~cm line \citep[e.g.][]{Seuss2016,Haynes2018,Hess2021,Roberts2021}, due to its relative proximity to the OH maser lines. The vast majority of these surveys have concentrated their efforts on surveying relatively large areas at low redshift in order to cover a large cosmic volume \citep{Darling2000,Darling2001,DarlingGio2002}. However, in order  to reach beyond the local Universe, high sensitivity is required, coupled with a relatively large bandwidth that allows a survey to cover a significant amount of cosmic volume by virtue of a deeper sampling in the radial direction, rather than broader areal coverage. One of the key facilities that is able to carry out this type of survey is MeerKAT \citep{MeerKAT,MeerKAT2}, which couples very high sensitivity with a wide bandwidth at both the L-band and Ultra-high-frequency (UHF) bands. Indeed, \cite{Glowacki2022} have already discovered a high-redshift ($z=0.52$) OHM using data from the MeerKAT telescope, as part of the Looking At the Distant Universe with the MeerKAT Array (LADUMA) Survey \citep{LADUMA}, and \cite{Combes2021} detected satellite-line absorption against the distant radio source PKS~1830-211 at $z = 0.89$ with known OH main-line absorption \citep{Chengalur1999}, as part of the MeerKAT Absorption Line Survey \citep[MALS; ][]{MALS}.

In this paper we report the discovery of the most distant OH megamaser found to date from an untargeted survey, using data from the MeerKAT radio telescope as part of the MeerKAT International Gigz-Hertz Tiered Extragalactic Exploration \citep[MIGHTEE; ][]{Jarvis2016} survey.  
In Section~\ref{sec:obs} we provide details of the MIGHTEE data and the calibration and imaging procedure used for creating the spectral line cubes. In Section~\ref{sec:OHM} we determine the properties of the OHM and the host galaxy in which it resides and in Section~\ref{sec:discuss} we discuss our results and summarise our conclusions.

Throughout the paper we assume $H_{\circ} = 67.7$\,km~s$^{-1}$~Mpc$^{-1}$, $\Omega_{\rm M} = 0.31$ and $\Omega_{\Lambda} = 0.69$ \citep{PlanckVI}.

\section{MIGHTEE Observations}\label{sec:obs}

The MIGHTEE survey is one of the Large Survey Projects currently being conducted by the MeerKAT radio telescope in South Africa. It is surveying approximately 20\,deg$^2$ over four of the most widely observed deep extragalactic fields accessible from the southern hemisphere. It is conducting the bulk of the survey using the L-band receiver, which covers the frequency range $856-1711$\,MHz. The Early Science data was taken with 4096 channels spanning the L-band, which has enabled a broad range of  science topics to be addressed using the radio continuum \citep[e.g.][]{Whittam2022, Hale2023}, spectral line \citep[e.g.][]{Maddox2021, Ponomareva2021, Ponomareva2023} and polarisation \citep[e.g.][]{Bockmann2023} data. Subsequent observations for the MIGHTEE survey were taken at the full resolution offered by MeerKAT after its correlator was upgraded, providing 32768 channels with a velocity resolution of $5.5$\,km\,s$^{-1}$ at 1420\,MHz.

The COSMOS field was observed by MeerKAT in 32k channel mode for a total of 15~$\times$~8 h tracks in a tightly-dithered mosaic that spans around 2~deg$^2$ at $\sim 1.4$\,GHz, resulting in 94.2~h of on-field integration. The target-only visibilities for each of these pointings were retrieved from the SARAO archive\footnote{\url{https://archive.sarao.ac.za}} at full spectral resolution using the KAT Data Access Library\footnote{\url{https://github.com/ska-sa/katdal}}, and with the Level-1 calibrations applied, as derived by the SARAO Science Data Processor. The MIGHTEE spectral line processing divides MeerKAT's L-band into three regions that are relatively free of radio frequency interference (RFI), namely 960--1150~MHz, 1290--1520~MHz, and 1610--1650~MHz. Each set of visibilities is split into these three sub-bands which are processed independently following Doppler correction to a barycentric reference frame. For the lowest frequency sub-band, the frequency domain is averaged by a factor of 4 to a resolution of 104.5~kHz, as we do not expect to detect low-velocity dispersion emission line sources in the low-frequency band, while the upper two sub-bands retain the full resolution.

Flagging of the visibilities is performed using the {\sc tricolour} package \citep{hugo2022}. Each sub-band is imaged using the {\sc wsclean} software \citep{offringa2014}, with a pointing-specific mask for the continuum sources derived from the existing deep MIGHTEE continuum images \citep{heywood2022}. The spectral clean component model is interpolated using the {\sc smops}\footnote{\url{https://github.com/Mulan-94/smops}} tool to provide smoothness in the spectral domain. Following inversion of this model into the visibility domain, a round of (phase+delay) self-calibration and simultaneous subtraction of the smoothed continuum model is performed using the {\sc cubical} \citep{kenyon2018} package. 

Each pointing is then imaged on a per-channel basis using {\sc wsclean} using three robustness (0.0, 0.5 and 1.0) parameters \citep{Briggs1995}, and deconvolution masks are constructed from the resulting image using a custom {\sc python} tool (Heywood et al. in prep). Imaging is repeated with deconvolution within the masked regions, and the resulting per-pointing cubes are homogenised to a common angular resolution per channel, using custom {\sc python} code as well as the {\sc pypher} package \citep{boucaud2016}. These homogenised images are primary beam corrected using the {\sc katbeam}\footnote{\url{https://github.com/ska-sa/katbeam}} library, and then linearly mosaicked assuming variance weighting using the {\sc montage}\footnote{\url{http://montage.ipac.caltech.edu/}} toolkit. A final process of image-plane continuum subtraction is performed using custom {\sc python} code along every sightline through the resulting cubes. Full details of the procedure outlined above and the custom methodology involved is provided in Heywood et al. (in prep.). We also note that the MIGHTEE data are taken in full polarisation mode, details of which can be found in Taylor et al. (in prep.).

\section{The OH megamaser J095903.22+025356.1}\label{sec:OHM}

\subsection{OH megamaser emission lines}
\begin{figure}
	\includegraphics[width=\columnwidth]{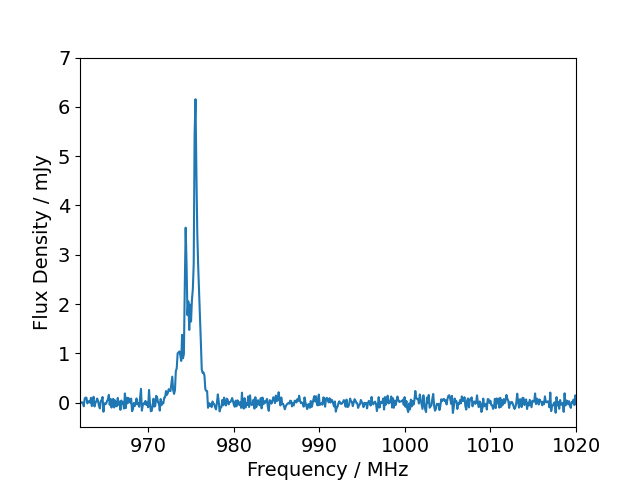}
    \caption{The observed-frame spectrum of the OH megamaser J095903.22+025356.1 from 960--1020\,MHz. The rms noise across the spectral range spans 70--80~$\mu$Jy per 104.5\,kHz channel across this range (see Heywood et al. in prep for further details).}
    \label{fig:ohm-spectrum-full}
\end{figure}

\begin{figure}
	\includegraphics[width=\columnwidth]{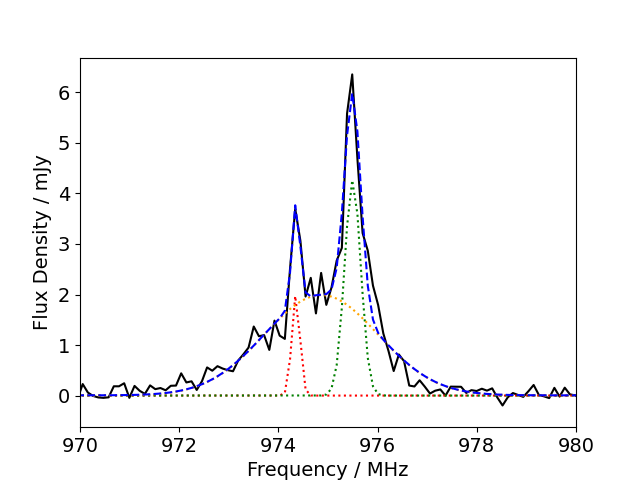}
    \caption{The observed-frame spectrum of the OH megamaser J095903.22+025356.1 (black solid line). The best fit ($\chi^2_{\rm red} = 0.6$) 3-Gaussian model (blue dashed line), includes broad emission from a redshifted 1667\,MHz line component (orange dotted line), along with narrow 1667\,MHz (green dotted line) and 1665\,MHz (red dotted line) emission lines.}
    \label{fig:ohm-spectrum-redshift}
\end{figure}

We visually inspected the low-frequency (960 - 1150\,MHz) robust-0.0 spectral-line cube, which has a spatial resolution of $10 \times 15$\,arcsec$^2$ and a median rms sensitivity of $\sim 75~\mu$Jy/channel (channel width of 104.5~kHz), over the COSMOS field to search for high-redshift emission line galaxies. We discovered a bright, unresolved source at RA = 09h\,59m\,03.22s Dec = 02d\,53m\,56.1s (J2000) and a frequency of $\nu = 975.31$\,MHz with a signal-to-noise (SNR) of $\sim 80$ at the peak of the line.  A second bright emission line centred at $\nu = 974.1$\,MHz was also observed at the same position on the sky (SNR = 51 at the peak of the line). The full spectrum, extracted over the restoring clean beam and normalised by the beam area, is shown in Fig.~\ref{fig:ohm-spectrum-full} and a zoom in on the 1665 and 1667\,MHz region is shown in Fig.~\ref{fig:ohm-spectrum-redshift}. These two lines correspond exactly to the rest-frame 1667\,MHz and 1665\,MHz main emission lines of the OH molecule at $z=0.7092$. Thus, this is the highest redshift OH main-line megamaser in emission discovered to date\footnote{We note that the highest-redshift OH megamaser in emission remains the tentative satellite-line detection of PKS1830-211 by \cite{Combes2021}, which was targeted as part of the MeerKAT Absorption Line Survey.}, eclipsing the previous record holder \citep[$z$~=~0.52;][]{Glowacki2022}, which was also discovered with MeerKAT.

As can be seen in Fig.~\ref{fig:ohm-spectrum-redshift}, the 1667\,MHz emission line is bright and relatively narrow, and the 1665\,MHz emission line also appears to have a narrow component. However, there also appears to be a broader underlying component around the 1665\,MHz line. We therefore initially fit the emission lines with three Gaussian profiles (two components for the 1665\,MHz line and a single component for the 1667\,MHz line), with the normalisation, width and redshift left as free parameters (assuming the redshift is the same for all three components). The resulting parameters are listed in Table~\ref{tab:ohm-properties} (top panel).

The ratio of the luminosities of the 1667 and 1665\,MHz lines  provides information on the physical conditions within the gas clouds from which the OH emission arises \citep[e.g.][]{DarlingGio2002}. Under the assumption of local thermodynamic equilibrium and optically thin lines, we would expect a line ratio of $1.8$, and considering only the narrow 1665\,MHz component {\bf ($3.4 \times 10^3$\,L$_{\odot}$)}, we measure a line ratio of 4.7, which is similar the ratio observed in other OHMs \citep[e.g.][]{McBride2013,Hess2021}. The observed line ratio is compatible with the models in which the main OH lines have non-negligible optical depths \citep[e.g.][]{LockettElitzur2008}.  

Unfortunately, we have no constraints on the satellite line at 1612~MHz as it falls below the lower end of the spectral coverage in our L-band data, and we find no evidence for emission at the redshifted 1720~MHz line at 1006.4~MHz. We estimate the limiting flux for the 1720~MHz line of $\sim 150\mu$Jy/beam (2$\sigma$) for the peak flux, and a corresponding integrated luminosity of $\log_{10}(L/L_{\odot}) = 3.08$ assuming a Gaussian line profile with FWHM = $120$~km~s$^{-1}$. This leads to an lower limit on the 1667/1720 ratio of $>3$, which is again within the range of line ratios exhibited in other lower-redshift OHM systems \citep{McBride2013}.

Given the very different line widths for the broad component and the 1667\,MHz line it is likely that the emission is coming from different regions, one with high velocity dispersion and a region at much lower velocity dispersion. Large gaseous outflows have been observed in infrared-luminous galaxies \citep[e.g.][]{Rupke2005,Spoon2013,Gowardhan2018}. Therefore, a possible second explanation for the broad component is that it is due to a high velocity outflow (or inflow) in the 1667\,MHz line. We therefore refit the emission line with an extra free parameter that allows a broad 1667\,MHz component to have a different redshift to the systemic redshift. The best fit parameters are given in Table~\ref{tab:ohm-properties} (bottom panel) and the fit is shown in Fig.~\ref{fig:ohm-spectrum-redshift}. In this case the inflow/outflow velocity of the 1667\,MHz line is $\sim 330$\,km\,s$^{-1}$ and provides a marginally better fit than the assumption of a broad 1665\,MHz line (reduced $\chi_{\rm red}^2 = 0.6$ compared to $\chi_{\rm red}^2 = 1.88$ for the broad 1665\,MHz component model). 
We note that more complex combinations of broad and narrow line components for the 1665 and 1667~MHz lines could replicate the observed spectrum. However, our current data do not support more complex models as the reduced $\chi_{\rm red}^2 = 0.6$ for the broad velocity outflow/inflow model, suggests we are already overfitting the observations. To explore this further, higher spatial resolution would be required, which requires longer baselines than currently available to MeerKAT. Thus, we are limited in terms of what we can say about the underlying emission line profiles, other than we find evidence for at least two distinct velocity dispersion components, and that we marginally prefer a model with a broad component that is either inflowing or outflowing with respect to the systemic velocity. We cannot differentiate inflow from outflow from the spectrum alone, due to the fact that we cannot determine whether the high-velocity component is in front of, or behind, the position of the narrow-line components, which we assume to trace the systemic redshift. However, the large width of the broad component is likely due to the large-scale motions of individual maser clouds \citep[e.g.][]{LockettElitzur2008} and suggests a system which is undergoing some level of disruption, possibly due to a major merger. This second model for the emission-line components reaffirm it as the brightest main-line OHM discovered thus far (see Table~\ref{tab:ohm-properties} for a summary of all the derived properties).

\begin{table}
	\centering
	\caption{({\em top}) Best-fit parameters for the 3-Gaussian fit to the emission lines, where the broad component is assumed to be the 1665~MHz line at the systemic redshift. ({\em bottom}) Best-fit parameters for the 3-Gaussian fit to the emission lines, where the broad component is assumed to be the 1667~MHz line that is redshifted with respect to the systemic velocity of the narrow lines. }
	\label{tab:ohm-properties}
	\begin{tabular}{lr} 
		\hline
		OHM properties with broad 1665 line &   \\
  \hline
  $z_{\rm spec}$ & $0.7092 \pm 0.0001$  \\
  $\log_{10}$($L_{\rm 1667}$/ L$_{\odot}$) & $4.20 \pm 0.02$  \\
    $\log_{10}$($L_{\rm 1665 n}$/ L$_{\odot}$) & $3.53 \pm 0.06$  \\
      $\log_{10}$($L_{\rm 1665 b}$/ L$_{\odot}$) & $4.46 \pm 0.03$ \\
  FWHM$_{1667}$ / km\,s$^{-1}$ & $172 \pm 6$ \\
   FWHM$_{1665\rm n}$ / km\,s$^{-1}$ & $72  \pm 8$ \\
   FWHM$_{1665\rm b}$ / km\,s$^{-1}$ & $917 \pm 48$ \\
   $\chi^2_{\rm red}$ & 1.88 \\
   	\hline
		OHM properties with broad redshifted 1667 line &   \\
  \hline
  $z_{\rm spec}$ & $0.7092 \pm 0.0001$  \\
  $\log_{10}$($L_{\rm 1667 n}$/ L$_{\odot}$) & $4.02 \pm 0.02$  \\
    $\log_{10}$($L_{\rm 1667 b}$/ L$_{\odot}$) & $4.54 \pm 0.02$ \\
    $\log_{10}$($L_{\rm 1665}$/ L$_{\odot}$) & $3.4 \pm 0.05$  \\
  FWHM$_{1667\rm n}$ / km\,s$^{-1}$ & $118 \pm 5$ \\
     FWHM$_{1667\rm b}$ / km\,s$^{-1}$ & $832 \pm 27$ \\
   FWHM$_{1665\rm n}$ / km\,s$^{-1}$ & $60  \pm 6$ \\
    Velocity of inflow / km\,s$^{-1}$ & $333 \pm 20$ \\
      $\chi^2_{\rm red}$ & 0.60 \\
  \hline
	\end{tabular}
\end{table}

We are able to constrain the pumping efficiency, $\eta_{\rm OH}$ of the maser emission using the following ratio:
\begin{equation}
\eta_{\rm OH} = \frac{L_{\rm OH} \times \Delta\nu_{IR}}{L_{IR} \times \Delta\nu_{\rm OH}} ,
\end{equation}
where $L_{\rm OH}$ is the luminosity of the 1667\,MHz OH line and$L_{\rm IR}$ is the infrared luminosity across the $53\mu$m line, which (supplemented by the 35, 80 and 120$\mu$m lines) can pump the OH molecule \citep[][]{Elitzur1982,LockettElitzur2008}. $\Delta\nu_{\rm OH}$ and $\Delta\nu_{\rm IR}$ are the widths of the OH and IR lines. Using Arp~220 as a local analogue \citep[e.g.][]{HeChen2004}, we assume pumping line widths $\sim 200$\,km\,s$^{-1}$ and equivalent widths $\sim 0.04\,\mu$m, and determine the rest-frame continuum emission at 53~$\mu$m from the SED fit in Fig~\ref{fig:sed} (we note that it is very close to the observed frame 100~$\mu$m measurement from {\em Herschel} so is well constrained). We find that the 53~$\mu$m line has an estimated luminosity of 5.2\,L$_{\odot}$, leading to an OH pumping efficiency of $\gg 100$ per cent for both the narrow and redshifted broad components of the 1667\,MHz line. This is not surprising given the extreme luminosity of the OHM, although as we discuss in Section~\ref{sec:lensing}, differential lensing may also lead to a much higher OHM luminosity relative to the host galaxy.
Given the 45~K dust  temperature inferred in Section~\ref{sec:host}, this estimate is in line with the finding of \cite{Kloecknerthesis}, suggesting that those OHM host galaxies with cooler ($<50$\,K) infrared emission exhibit greater efficiency ($\gg 1$ per cent), and that the maser emission is saturated. Such saturated maser emission is thought to arise from  high-density, optically-thick regions ($>10^5$~cm$^{-3}$) \citep{Elitzur1982,Darling2007}, which are also some of the strongest maser sources.

\subsection{Host galaxy properties}\label{sec:host}

\begin{figure*}
	\includegraphics[scale=0.28]{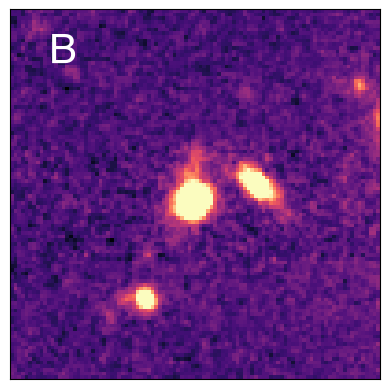}
 \includegraphics[scale=0.28]{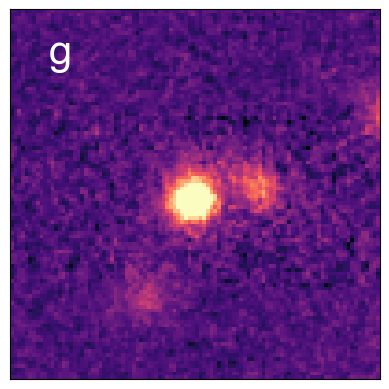}
 \includegraphics[scale=0.28]{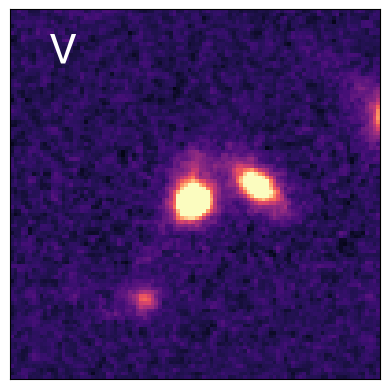}
 \includegraphics[scale=0.28]{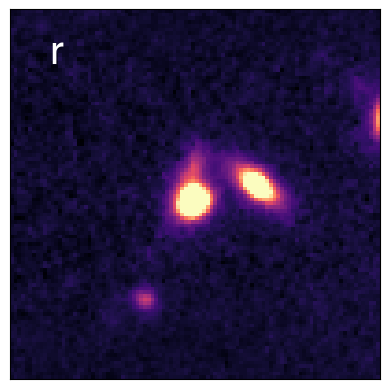}
 \includegraphics[scale=0.28]{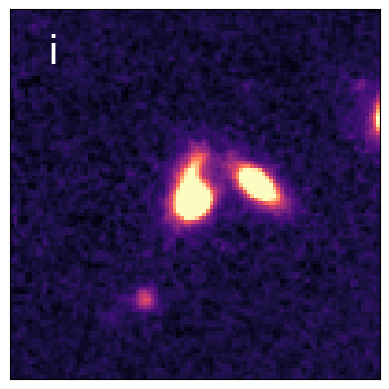}
 \includegraphics[scale=0.28]{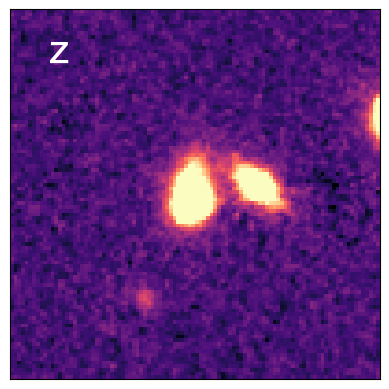}
    \caption{Postage stamps of the OHM host galaxy in $B$,$g$,$V$,$r$,$i$ and $z$-band filters from Subaru. The galaxy $\sim 2.6$~arcsec to the west has a photometric redshift of $z_{\rm phot} = 0.35$ or $0.43$ (see text) and is not associated with the OHM host galaxy. Note that the different filters have different depths and the colour scale is chosen to bring out the key features. The flux-density of the OHM host galaxy for each filter is given in Table~\ref{tab:photometry}.}
    \label{fig:postage}
\end{figure*}

\begin{figure}
	\includegraphics[width=\columnwidth]{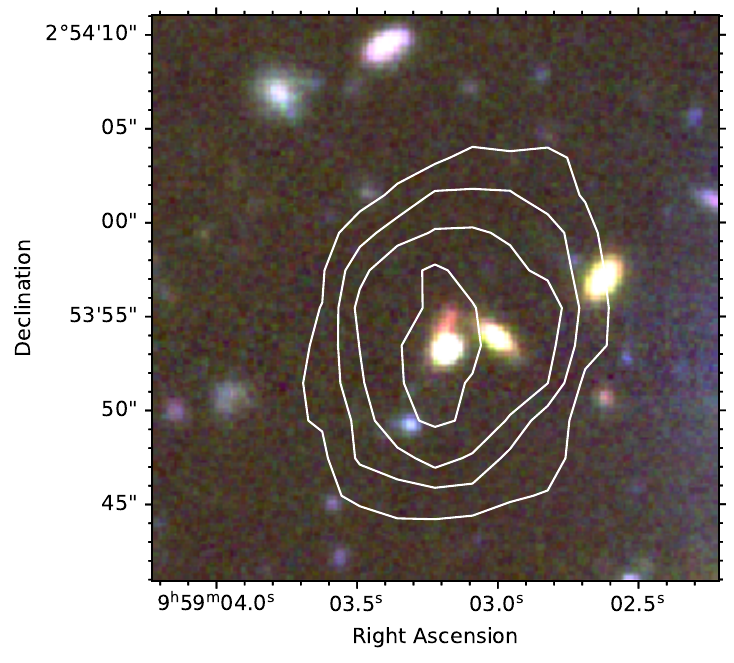}
    \caption{3-colour ($B$, $V$ and $i$) image of the OHM host galaxy with the OHM emission denoted by the contours (40, 70, 100 and 140\,Jy\,Hz). The OHM is unresolved at the spatial resolution of our data and as such the contours represent the restoring beam.}
    \label{fig:ohm-contours}
\end{figure}

The COSMOS field has been widely observed across the full electromagnetic spectrum, we therefore have a wealth of multi-wavelength data with which to measure the properties of the OH megamaser host galaxy. We have identified an optical source at the position of the OH megamaser using the COSMOS2020 catalogue \citep{Weaver2022}. 
In Fig.~\ref{fig:postage} we show the $B$, $g$, $V$, $r$, $i$ and $z$-band imaging around the OHM from the Subaru telescope, and in Fig.~\ref{fig:ohm-contours} we show the emission from the OHM overlaid on a 3-colour optical image. The host galaxy is clear alongside a  galaxy 2.6~arcsec away (but at a lower redshift - see Sec.~\ref{sec:lensing}). The OHM host galaxy itself appears to have a stream of emission across all visible bands to the north, which may be indicative of an interaction, i.e. a tidal tail from a merger event. However, the ground-based data are not high enough resolution to disentangle this emission and confirm whether it is due to a merger. Unfortunately, the OHM falls outside the {\em HST} coverage of the COSMOS field. The other possibility is that this elongated emission is due to gravitational lensing by the nearby galaxy in projection and lies at a lower redshift, and we return to this possibility in Section~\ref{sec:lensing}.

The available $u-$band through to {\em Spitzer}/IRAC photometry is presented in Table~\ref{tab:photometry}, supplemented with mid-infrared data from the {\em Wide-field Infrared Survey Explorer} ({\em WISE}) and the {\em Spitzer Space Telescope} and far-infrared data from the {\em Herschel Space Observatory} using the Herschel Extragalactic Legacy Project (HELP) database\footnote{https://hedam.lam.fr/HELP/} \citep{HELP}.  We are also able to measure the radio continuum flux from both the MIGHTEE data itself, with an effective frequency of 1.2GHz at the position of the OHM \citep{heywood2022}, and using the VLA 3GHz survey of the COSMOS field \citep{Smolcic2017}\footnote{We note that the VLA 3GHz data may resolve out some emission from this host galaxy, see \citep[e.g.][]{Hale2023}.}.

\begin{figure*}
	\includegraphics[width=\textwidth]{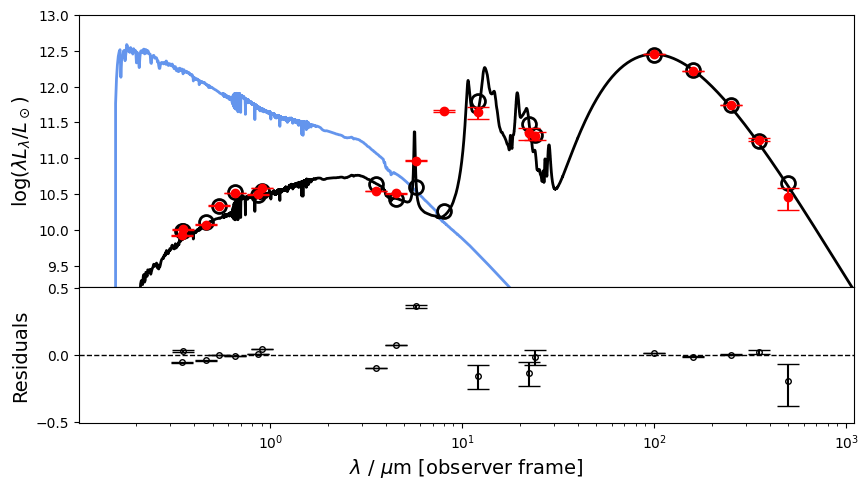}
    \caption{Spectral energy distribution of the OHM host galaxy with the best fit to the combined optical, near-infrared and far-infrared SED, derived using {\sc MagPhys} (black line). The blue line represents the intrinsic stellar spectrum that has been reddened with $A_{\rm V} = 2.94$ in order to fit both the optical and near-infrared data and the reprocessed dust emission at mid- to far-infrared wavelengths. The red solid circles are the observed photometry, and the open black circles denote the predicted photometry from the best-fit model.
    The best-fit parameters are listed in Table~\ref{tab:ohm-host-properties}. The residuals of the fit with respect to the data are presented in the lower panel.The point at $8\mu$m lies $\sim 1.6$~dex above the best fit SED and is therefore not shown in the the lower panel.}
    \label{fig:sed}
\end{figure*}

In order to determine the properties of the host galaxy, we use {\sc MagPhys} \citep{daCunha2008} to perform a fit to the spectral energy distribution (SED), in which the energy absorbed in the UV part of the spectrum is balanced with the energy reemitted at far-infrared wavelengths. The best-fit SED along with the measured photometry is shown in Fig.~\ref{fig:sed} and the best fit parameters are presented in Table~\ref{tab:ohm-host-properties} after fixing the redshift to $z= 0.7092$, assuming a Chabrier Initial Mass Function \citep{chabrier2003}. We are able to obtain a very good fit to the vast majority of the photometric data (the exception being the measurement at 8$\mu$m, which we return to below). The key derived properties suggest a rapidly star-forming galaxy with a star-formation rate (SFR) of $371 \pm 20$\,M$_{\odot}$\,yr$^{-1}$ and a total stellar mass of $\log_{10}(M_{\star}/$M$_{\odot}) = 10.5\pm 0.2$, which together means that it lies approximately 1.5\,dex above the star-forming galaxy main sequence at this redshift \cite[e.g.][]{Whitaker2014,Johnston2015}. The stellar mass derived from {\sc MagPhys} is also consistent with that determined in the COSMOS2020 catalogue, which reports a total stellar mass of $\log_{10}(M_{\star}/$M$_{\odot}) = 10.31 \pm 0.04$. However, we find a much  higher SFR compared to the value in the COSMOS2020 catalogue ($\approx 50$\,M$_{\odot}$\,yr$^{-1}$), which is unsurprising given the level of dust extinction from the model fit ($A_V = 2.94$) and  the significant amount of obscured star formation evidenced from the high far-infrared luminosity of $L_{\rm IR} = 3.6\times 10^{12}$\,L$_{\odot}$, which confirms  the host galaxy as an Ultra-Luminous Infrared Galaxy (ULIRG).

\begin{figure}
    \includegraphics[width=\columnwidth]{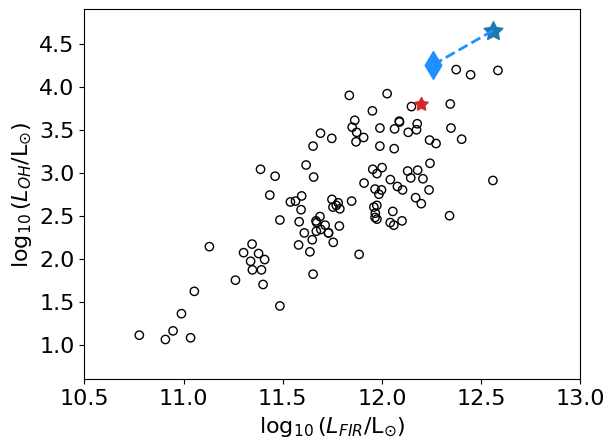}
    \caption{The far-infrared -- OH megamaser luminosity relation for the sample of OHMs from Roberts \& Darling (in prep) (open circles), the OHM from \citet{Glowacki2022} and the OHM: J95903.22+025356.1, presented in this paper (blue star). The OHM luminosity shown is the combination of the broad- and narrow-1667\,MHz emission line luminosities. The dashed line towards the light-blue diamond denotes where the OHM and host galaxy would lie if the OHM was magnified by a factor of 2.5 and the host by a factor of 2 (the two most likely lensing magnifications).}
    \label{fig:LFIR_LOH}
\end{figure}

Such galaxy properties are similar to the host galaxies of OH megamasers in the local Universe, where there is a strong correlation between the OH megamaser luminosity and the far-infrared luminosity \citep[e.g.][]{DarlingGio2002,DarlingGio2006,Zhang2014}. To highlight the properties of this OHM, in Fig.~\ref{fig:LFIR_LOH} we show the OH luminosity against far-infrared luminosity for a complete sample of OHM galaxies at $z<0.5$ from the compilation Roberts \& Darling (in prep.) and the OHM at $z=0.52$ from \cite{Glowacki2022}. The $z=0.7092$ OHM discussed in this paper is clearly the brightest OHM thus far discovered, but the far-infrared luminosity is also very high, such that it lies along the known correlation. Thus, although very luminous, its ratio between far-infrared and OH luminosities is not significantly different from the low-redshift populations.

As mentioned earlier, one interesting detail in the SED is that the measured flux in  IRAC Channel 4 at 8$\mu$m is around 1.5~dex higher than the best-fit SED. This is the only measurement that does not agree within the uncertainties with the best-fit SED, and is such an extreme outlier it warrants discussion. We note that the galaxy is also very bright in the $W3$ filter, with a very red colour between the $W2$ and $W3$ bands, thus confirmiung that the 8$\mu$m emission is unlikely to be a spurious measurement.
One possible explanation is that the total SED could be a combination of multiple galaxies within the PSF of the 8$\mu$m data. However, the only galaxy that is nearby is $\sim 2.6$\,arcsec away and does not appear to show a strong increase in flux from the 3.6 to $8\mu$m bands in the COSMOS2020 catalogue, where $S_{3.6} = 9.86\mu$Jy and $S_{8} = 22.5\mu$Jy (compared to the flux from the maser host galaxy of $S_{8} = 2076\mu$Jy). We therefore rule out contamination from the projected nearby galaxy. Moreover, as can be seen in Fig.~\ref{fig:sed}, the other mid-infrared bands all require significant emission from polycyclic aromatic hydrocarbons (PAHs); in particular the rest-frame 3.3$\mu$m feature is needed to fit the 5.8$\mu$m photometry. The presence of strong PAHs is not unusual in OHM host galaxies \citep[e.g.][]{Willett2011a}, again suggesting that the OHM presented here is not anomalous and has very similar properties to the vast majority of other OHM host galaxies.

Another possible explanation for the excess emission in the 8$\mu$m band could be hot dust emission from a torus around an accreting active galactic nucleus (AGN). We therefore use the {\sc Cigale} SED fitting code \citep{Bouquien2019} to attempt to fit a composite AGN+galaxy SED. We adopt parameters similar to those used in studies of radio-selected AGN \citep{Zhu2023,Best2023} and more broadly selected galaxy samples \citep{Zou2022}, which incorporate two models for AGN emission from both the accretion disk and the obscuring torus \citep{Fritz2006,Stalevski2012,Stalevski2016}, alongside stellar population synthesis models from \cite{bc03}. However, there is no combination that can reproduce the high luminosity within the 8$\mu$m band, whilst also still adequately fitting the range of other multi-wavelength data. We are likewise not aware of any spectral feature that could be boosting the $8\,{\rm \mu m}$ flux density at the redshift of the OHM system.

In order to check whether there are any other signatures from an AGN component, we make use of the radio continuum imaging data from both the MIGHTEE data and the VLA COSMOS 3\,GHz survey \citep{Smolcic2017}. We measure a flux density at 1.2\,GHz of $S_{1.2\rm GHz} = 250 \pm 9$\,$\mu$Jy, which corresponds to a SFR = $342 \pm 68$\,M$_{\odot}$~yr$^{-1}$; the uncertainty encompasses the systematic uncertainty of using different conversions from 1.4\,GHz to a SFR from \cite{Bell2003}, \cite{Delhaize2017} and \cite{Delvecchio2021}. Thus, the SFR using the best-fit {\sc MagPhys} SED is completely consistent with the SFR estimated from the radio continuum. Moreover, the radio continuum emission is unresolved at the highest resolution of both the MIGHTEE data ($\sim 5$\,arcsec) and the VLA COSMOS data ($\sim 0.7$\,arcsec). Taken together, the radio emission does not provide any indication that there may be a contribution to the SED from an AGN\footnote{We note that the position of the OH megamaser is not covered by Chandra X-ray survey over the COSMOS field \citep{Civano2016}.}.
Therefore, although the SED fit and the radio continuum measurements are consistent, the excess emission at mid-infrared wavelengths remains a puzzle and will require much higher resolution mid-infrared data to resolve, e.g. from {\em JWST}. 

We can also use the measurements from {\em WISE} and {\em Spitzer}/IRAC to determine where the host galaxy of the megamaser resides in colour space compared to the broader classifications of galaxy populations. Using the colour-colour plots from \cite{Roberts2021}, who investigate where the host galaxies of masers should reside compared to those of H{\sc i} galaxies, we find that the host galaxy of OHM J095903.22+025356.1 has {\em WISE} and IRAC colours fully consistent with the local OHM host galaxy population. However, we note it has a relatively red $W2-W3 = 6.15$\,mag, meaning it lies at the extreme red end of the ULIRG population in the {\em WISE} colour-colour diagram of \cite{Jarrett2011}.  The extreme $W2-W3$ colour is partly due to the fact that the $W2$ and $W3$ bands are sampling the rest-frame $\sim 2.7$ and $\sim 7.1\mu$m emission, and that at these rest wavelengths, the PAH emission contributes significantly to the measured photometry, particularly in the W3 filter. 
However, we note the $W1-W2 = 0.714$ colour is very much in the middle of the expected range for ULIRGs and starburst galaxies and lies within the colour range expected for OHM host galaxies \citep[see e.g.][]{Roberts2021, Glowacki2022}, with this emission predominantly arising from old stellar populations, rather than warm dust and PAH emission.

\begin{table}
	\centering
	\caption{Measured photometry for the OHM host galaxy from the COSMOS2020 catalogue complemented with mid- and far-infrared data from {\em WISE}, {\em Spitzer} and {\em Herschel} using products from the {\em Herschel} Extragalactic Legacy Project \citep[HELP; ][]{HELP} and radio data from MIGHTEE and the VLA 3GHz survey. {\em WISE} filters with a $\dagger$ superscipt are not used in the SED fitting due to better similar wavelength data from {\em Spitzer}. }
	\label{tab:photometry}
	\begin{tabular}{lcrr} 
		\hline
		Filter & Effective Wavelength & $S_{\nu}$ & $\sigma_{S_{\nu}}$ \\
     & ($\mu$m) & ($\mu$Jy) &  ($\mu$Jy) \\
     \hline
		$u$ & 0.346 & 1.65 & 0.02\\
		$u^*$ & 0.350 & 2.03 & 0.03\\
		$g$ & 0.460 & 3.11 & 0.02\\
        $r$ & 0.538 & 6.68 & 0.03\\
        $i$ & 0.652 & 12.16 & 0.03\\
        $z$ & 0.866 & 15.36 & 0.05\\
        $y$ & 0.906 & 19.68 & 0.08 \\
        IRAC1 & 3.56 & 71.1 & 0.2 \\
        IRAC2 & 4.51 & 82.8 & 0.2 \\
        IRAC3 & 5.76 & 304 & 7 \\
        IRAC4 & 8.00 & 2076 & 80 \\
        W1$^{\dagger}$ & 3.37 & 72.0 & 14.4\\
        W2$^{\dagger}$ & 4.62 & 99.2 & 19.8 \\
        W3 & 12.08 & 3003 & 600  \\
        W4 &   22.19 & 2813 & 560 \\
        MIPS & 24 & 2785 & 347 \\
        PACS100 & 100 & 164496 & 435 \\
        PACS160 & 160 & 151030 & 540 \\
        SPIRE250 & 250 & 78641 & 834 \\
        SPIRE350 & 350 & 36268 & 1396 \\
        SPIRE500 & 500 & 8146 & 2807 \\
        \hline
        \hline
        Radio Band & Effective Frequency & $S_{\nu}$ & $\sigma_{S_{\nu}}$\\
        &  & ($\mu$Jy) &  ($\mu$Jy) \\
        \hline
        L-band & 1.19~GHz & 250 & 9 \\
        S-band & 3.0~GHz & 118 & 7 \\
		\hline
	\end{tabular}
\end{table}

\begin{table}
	\centering
	\caption{Best fit parameters for the SED from {\sc MagPhys}.
 SFR in the radio is calculated assuming the measured spectral index between 1.2 and 3~GHz, and the flux at 3~GHz from VLA COSMOS.}
	\label{tab:ohm-host-properties}
	\begin{tabular}{lr} 
		\hline
    Host galaxy properties & \\
    \hline
    $z_{\rm phot}$ & $0.706^{+0.09}_{-0.08}$\\
		 $\log_{10}$($M_{\star}$ / M$_{\odot}$) & $10.47\pm 0.2$\\
		SFR$_{\rm SED}$ (M$_{\odot}$\,yr$^{-1}$) & $371 \pm 20$ \\
  SFR$_{\rm Radio}$ (M$_{\odot}$\,yr$^{-1}$) & $342 \pm  68$  \\
		 $\log_{10}$(sSFR$_{\rm SED}$/ yr$^{-1}$) & $-7.92 \pm 0.6$ \\
   $\log_{10}$($L_{\rm dust}$ / L$_{\odot}$) & $12.56 \pm 0.02$ \\
   $T_{\rm dust} /$ K & $45 \pm 2$\\
   $\log_{10}$($M_{\rm dust}$ / M$_{\odot}$) & 8.27 $\pm 0.04$ \\
   $A_{V}$ & $2.94 \pm 0.02$ \\
   $\log_{10}$($L_{1.4\rm GHz}$/ W\,Hz$^{-1}$) & $23.90 \pm 0.04$  \\
		\hline
	\end{tabular}
\end{table}

\subsection{Lensing}\label{sec:lensing}

\begin{figure*}
    \centering
    \includegraphics[width=\textwidth]{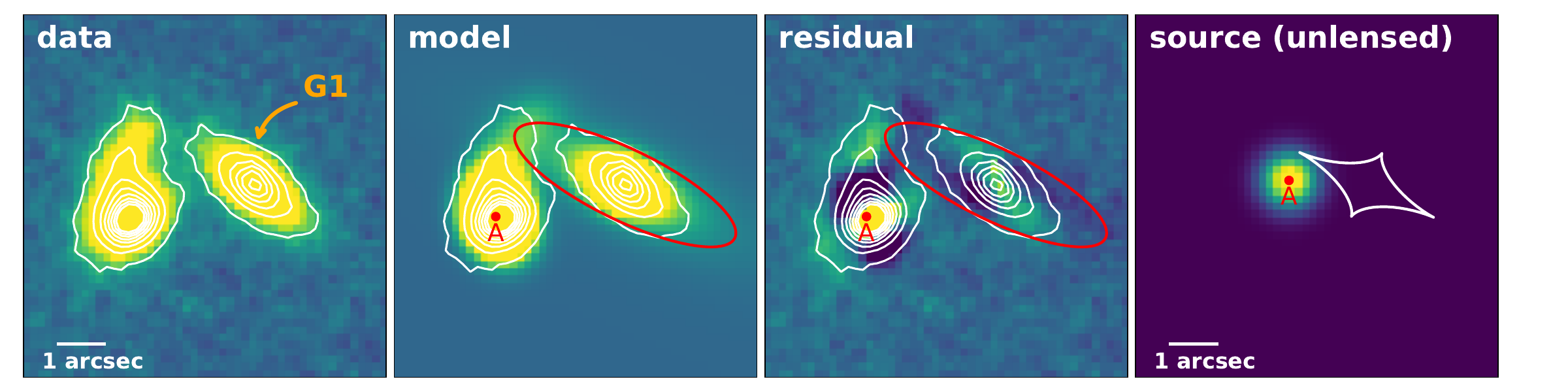}
    \caption{Left: Data from the Subaru {\sl i} band image and overlaid white self-contours, with galaxy G1 labelled. Middle left: Median posterior model from the single lens model convolved with the PSF. The critical curve is overlaid in red and the median image-plane position of the source centroid is labeled `A'. A point source at this location has a magnification of $\mu_{\rm pt,A} = 20.4$, while the entire source has a magnification $\mu_{\rm tot} = 1.8$. Middle right: residual image (data - model) with data contours, critical curve, and point A indicated. Right: Source plane representation of the lensing with a circular Gaussian of radius $R_{\rm src} = 0.25$~arcsec. The caustic curve is shown in white and the image stretch ranges from the zero to the peak of the source-plane Gaussian. As can be seen, the centroid `A' lies very near the cusp caustic. The scale-bar corresponds to the source plane.  }
    \label{fig:datamodelresid}
\end{figure*}

Given the very high luminosity of the OH emission line and the stream of emission to the north of the OHM host galaxy (Fig.~\ref{fig:postage}), it is worth assessing whether the system could be gravitationally lensed. The host galaxy mass and star-formation rate are not extreme however, suggesting that the host is unlikely to be significantly lensed. If the OHM originates from a very compact region within the host galaxy, then differential lensing could still provide significant magnification to the emission lines. There is a galaxy $2.6$~arcsec away from the OHM host galaxy (the galaxy lying directly to the west in Figs. \ref{fig:postage}, \ref{fig:ohm-contours} and \ref{fig:datamodelresid}, referred to as G1), which is present in the COSMOS2020 catalogue. It has a best-fit photometric redshift of $z_{\rm phot} = 0.43^{+0.02}_{-0.03}$ and stellar mass of $\log_{10}( M_{\star} / $M$_{\odot}) = 9.56$ using {\sc Le Phare} \citep{Ilbert2006} and $z_{\rm phot} = 0.35\pm0.01$ and $\log_{10}(M_{\star} / $M$_{\odot}) = 9.75$ derived using the {\sc Eazy} \citep{eazy}.

The Einstein radius for a lens of this mass and redshift (similar for both photometric redshifts) is $\sim 1.1$~arcsec, assuming  a halo mass of $M_{\rm halo} = 3.2\times 10^{11}$\,M$_{\odot}$, estimated using an empirically derived stellar-mass to halo-mass ratio \citep{Behroozi2010}. 

The primary aim of our lens modelling is to estimate a plausible magnification factor range of the OH emission, which is spatially unresolved with the MeerKAT $10 \times 15$~arcsec beam. Our objective is to robustly constrain a macro lens model, rather than precise non-parametric reconstruction of the stellar light distribution. This approach is taken for the following reasons: (a) the $\sim 0.7$~arcsec seeing of the ground-based data limits the inference possible for this high ellipticity, small Einstein ring foreground lens; (b) the lack of constraints from unambiguously identified multiply lensed images; and (c) the expectation that the source has an intrinsically complex, asymmetric morphology that is typical of ULIRGS \citep[e.g.][]{Clements1996,Veilleux2002,Yuan2010,Larson2016}.

We perform our lens modelling analysis with the {\sc lenstronomy}\footnote{https://github.com/lenstronomy/lenstronomy} package \citep{Birrer2018,Birrer2021}, enabling a Bayesian approach to the lens model parameter estimation. We employ the Particle Swarm Optimization non-linear fitting routine to provide the starting point for the Markov Chain Monte Carlo sampler, as described in \citet{Birrer2015}.  

We explore two lens models: the first is the `single lens model' with the closest galaxy (G1) as the only lens, with a projected separation of 2.4 arcsec, the second model, the `two lens model', includes a more distant galaxy (G2) which is 9.1~arcsec away to the north west (and can be seen in Fig~\ref{fig:ohm-contours}), and may perturb both the convergence and shear of the lensing system. G2 is also listed in the COSOMOS2020 catalogues and has a photometric redshift of $z_{\rm phot} = 0.35$, with stellar mass $M_{\star} = 1.9 \times 10^{10}$\,M$_{\odot}$ ($M_{\rm halo} = 8.8 \times 10^{11}$\,M$_{\odot}$), which has an Einstein radius of $\theta_{\rm E} =1.8$~arcsec.

Both models assume Single Isothermal Ellipsoid (SIE) mass density profiles for the foreground lens(es) and lens redshift(s) of $z= 0.35$. Both models also assume lens light follows mass (i.e. the lens light and mass distributions have co-located centroids and matching ellipticity and position angle). We first fit the G1 galaxy light profile without considering any lensing (i.e. $\theta_{\rm E} = 0$), deriving a S\'ersic index of $n_{\rm s} = 3.74$. We find that MCMC convergence for all subsequent lens modelling (i.e. $\theta_{\rm E} > 0$) requires that we fix the main lens light profile S\'ersic index to this value. We note that if there is a counter-image blended with the lens light, this approach may bias the lens light profile, and, hence the derived lens model. We test the sensitivity to this assumption by fixing the S\'ersic index at a few values in the range $3 < n_{\rm s} < 4$ and find consistent lens models within the statistical uncertainties. For the two lens model, the S\'ersic index of G2 is a free parameter, with a median posterior value of $n_{\rm s} = 2.31\pm0.02$. Both lens models assume a circular Gaussian source, as we find large degeneracies in any model that deviates from this simplistic assumption. The resultant residuals are large, as one would expect, but given our objective of macro lens model constraints with the available data, we find this an appropriate level of model complexity.

Using the single lens model we find a lens Einstein radius for G1 of $\theta_{\rm E} = 1.32 \pm 0.16$~arcsec, while with the two lens model we find a slightly lower value of $\theta = 1.16 \pm 0.16$ arcsec. The uncertainty is dominated by the systematics of lens model selection, and we assign an indicative uncertainty as the difference between the two derived values. Both of the Einstein radius values are comparable to that derived from the estimated halo mass, $\theta_{\rm E,halo} = 1.08$~arcsec. This supports both the lensing hypothesis and the derived macro lens models. In Fig.~\ref{fig:datamodelresid}, we show the results of the single lens model.

\begin{figure}
    \centering
    \includegraphics[width=\columnwidth]{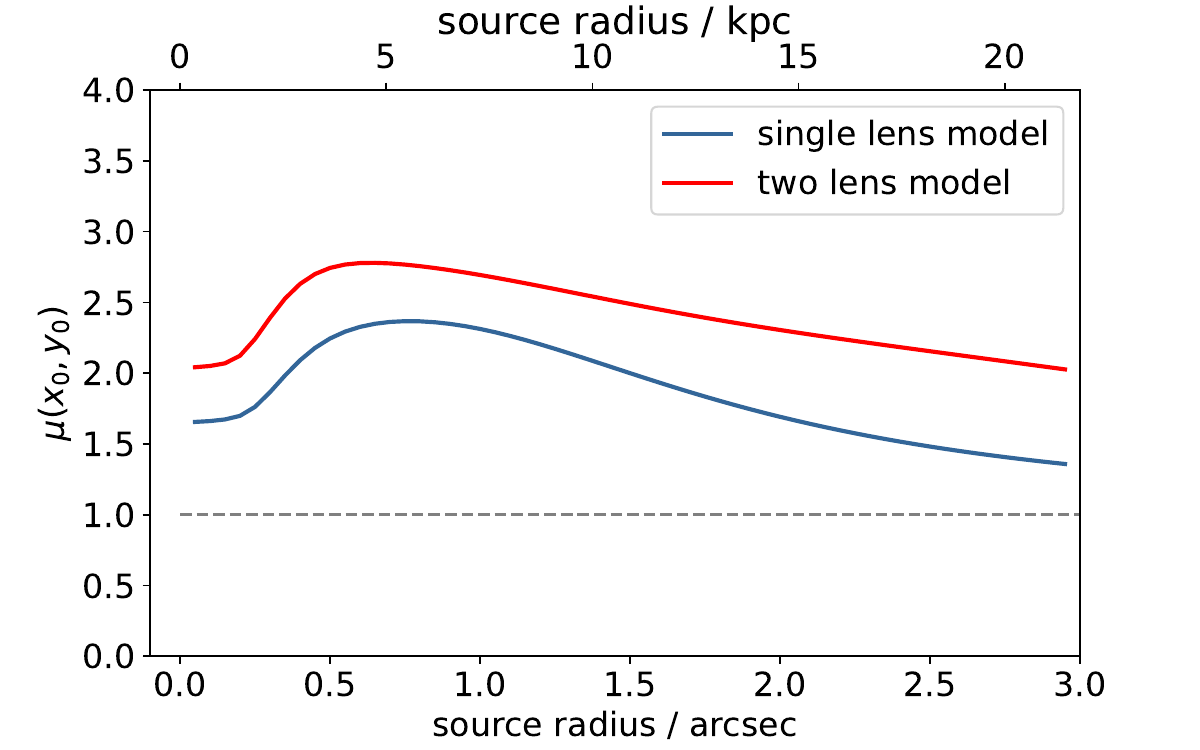}
    \caption{Magnification, $\mu$, as a function of source-plane radius for a circular Gaussian located at the median value of the centroid posterior probability density functions, ($x_0,y_0$). The blue and red curves correspond to the single lens model and two lens model, respectively, with the latter showing systematically higher magnification. The `bump' seen at 0.5-1 arcsec is a result of the source size enveloping the entire caustic, resulting in higher total magnification. }
    \label{fig:lensingmag}
\end{figure}

In Fig.~\ref{fig:lensingmag} we show the lensing magnification expected for a source at the redshift of the OHM as a function of angular and physical source size for both lens models. This shows that if the OHM is relatively compact ($\lesssim  1$~arcsec), then the magnification could be of the order of $\mu \sim 2.5-3.5 \pm1.0$, where the uncertainty is assumed to be dominated by the systematic uncertainty between the two models, which has an average of 36 per cent for radii between 0.05 and 3 arcsec. 
Even larger magnifications ($\mu > 5$) are possible if the OHM is very compact ($\leq 0.1$\,arcsec) and the OHM happens to lie closer to the caustic than the centroid of the optical emission, as could easily happen in a complex merging system.
The physical size of OHM emitting regions observed with Very Long Baseline Interferometry (VLBI) show them to be compact ($1-100$\,pc in size) \citep{Chapman1990,Lonsdale1994,Diamond1999,Polatidis2000,Rovilos2003,Kloeckner2003,Pihlstrom2005,Momjian2006,Baan2023}, thus a higher degree of lensing is plausible.

 However, even with this lensing magnification (and assuming no lensing of the more extended host galaxy) it would still remain consistent with the $L_{\rm FIR} - L_{\rm OHM}$ relation (see lower point in Fig~\ref{fig:LFIR_LOH}). 
This differential lensing of compact components near the host galaxy nucleus in a near cusp-cautic lensing configuration is very similar to that seen in \citet{Deane2013ii,Deane2023iii}, where the compact AGN core had an order of magnitude higher magnification factor than the stellar and cold molecular gas emission.  
We note that the lens model cannot account for all of the emission to the north of the OHM host galaxy (see residual in Fig.~\ref{fig:datamodelresid}).  The most likely reason for this emission is due to an interaction, combined with the limitations of a simple circular Gaussian source model.

Given the likelihood of lensing of a compact OHM towards the centre of the host galaxy, it is worth revisiting the excess emission around $8~\mu$m. A very compact region of hot dust could be both very bright around the $8~\mu$m region and, given the lensing probability, magnified in the same way as the OHM (or more, depending on how compact the region is and where it reside with respect to the caustic). If we assume that the hottest dust is emitted from the inner parts of a putative torus around a supermassive black hole, which is the most likely source of such compact hot dust, then the typical radius is $1-10$\,pc \citep{Tristram2009,Tristram2011}, similar in size to OHM emission regions. Thus, differential lensing of a very compact hot dusty region could magnify the mid-infrared emission by a similar amount to the OHM emission, and possibly more depending on the geometry and distribution of the dust. Again, it is very difficult to determine the actual amount of magnification with the resolution of the ground-based data, and this becomes even more difficult at the longer wavelengths, where we are currently reliant on {\em WISE} and {\em Spitzer} data which have much poorer resolution than the visible-wavelength data. However, we note that such an additional source of emission at mid-infrared wavelengths would also likely remove the requirement to have such strong PAH emission as currently required in the SED fit shown in Fig.~\ref{fig:sed} and reduce the large residuals (shown in the lower panel of Fig.~\ref{fig:sed}) at these mid-infrared wavelengths. This is because a significant fraction of the mid-infrared emission could originate from the hot dusty torus, and would therefore be magnified. The degree of magnification would depend on both the temperature distribution and the spatial distribution of the dust.
It is clear that there are many uncertainties around this system, both due to the possible lensing and the likelihood of an ongoing merger, and the current data preclude us from investigating this further. Thus, higher-resolution multi-wavelength observations with {\em JWST} are needed to resolve the question of lensing in this object, and to understand the system more fully.

\section{Conclusions}\label{sec:discuss}

We have presented details of the most distant known OH megamaser in the main OH lines discovered thus far at a redshift of $z=0.7092$. Analysis of the spectrum of the OHM suggests that there are two velocity components that produce the overall line profile, one in which the OHM gas has a relatively low velocity dispersion (FWHM$ ~\sim 100$\,km~s$^{-1}$) and from which we observe both the 1665 and 1667\,MHz lines from the OHM molecule. The second component is significantly broader, with a full-width half maximum of 832\,km~s$^{-1}$ and is redshifted with a velocity of 333\,km~s$^{-1}$ relative to the narrow OHM lines. This latter component could be either an inflow or outflow (depending on whether it lies in front of or behind the region responsible for the narrow emission).
However, the optical imaging shows what appears to be a tidal-tail feature similar to what is expected from a gas-rich merger, that cannot be explained by gravitational lensing. We therefore infer that the broad velocity component is dynamically separate from the galaxy from which the narrower OHM emission lines arise. The merger process also naturally produces sufficiently high infrared luminosity to pump the OH molecules and produce the very high luminosity of the OHM line ($L_{1667} >10^4$\,L$_{\odot}$).

We fit the host galaxy spectral energy distribution and find that the host is consistent with the  properties of OHM  galaxies in the local Universe, i.e. having a high star-formation rate (SFR $\sim 350$\,M$_{\odot}$~yr$^{-1}$) and high dust oscuration. The measured SFR and stellar mass of the host suggest that it lies 1.5~dex above the star-forming galaxy main sequence at this redshift. The close agreement between star-formation rate estimated from the SED fitting and the radio continuum measured at both 1.2 and 3\,GHz suggests that there is no evidence for AGN-related emission from the host. Indeed, we do not find a combined galaxy+AGN SED model that could reproduce the measured photometry. 
However, the OHM host galaxy lies in close proximity to a galaxy 2.6~arcsec away in projection (at $z\sim 0.35$) which may be gravitationally lensing the OHM emission (and possibly compact hot dust emission from an AGN torus for example, around observed wavelengths of $\sim 8\mu$m). 
The lensing magnification depends on the exact size and location of the OHM emission region, with magnification factors $\mu \sim 2.5$ for emission regions on the $\sim 1$\,arcsec scale of the host galaxy, but potentially higher ($\mu > 5$) if the emitting regions are similar in size to OHM emission regions in local galaxies and displaced towards the lensing caustic. However, the ground based data preclude us from making stronger statements on the level of magnification of the OHM and the hot dust emission.
We note that the gravitational lensing would produce similar magnification factors to both the optical and infrared data that trace the star-formation and the near-infrared data that traces the stellar mass, if they arise on similar spatial scales. Thus the interpretation of a starburst galaxy still holds in the event of significant gravitational lensing.

The discovery of this OHM, along with another high-redshift 
megamaser recently discovered by MeerKAT \citep{Glowacki2022}, points towards a plausible new window on the obscured galaxy population at high redshift, providing an observational realisation of previous proposals \cite{Briggs1998,Townsend2001, DarlingGio2002}. The fact that two new megamasers have been found at $z>0.5$ within just two of the MeerKAT fields thus far analysed suggests that many more will be discovered with the full MIGHTEE survey, which has a factor of $\sim 10$ more area than analysed to date, albeit at slightly lower sensitivity. Indeed, assuming approximately one OHM per MeerKAT primary beam at $\sim 1$\,GHz, using the OHM discussed in this paper and that found by \cite{Glowacki2022} as a guide, then we should expect to find around 10-20 more high-redshift ($0.45 \lesssim z \lesssim 0.8$) OHMs over the full MIGHTEE survey area (Jewell et al. in prep.).
This effort should lead to new constraints on the evolution of the most obscured systems in the Universe and possibly an independent measure of the gas-rich galaxy merger rate \citep[e.g.][]{Briggs1998}. These new discoveries will come from spectral line surveys, meaning that the uncertainty in photometric redshifts, particularly for these very obscured systems, is negated \cite[although some uncertainty may remain around confusion between OH and H{\sc i} lines; ][]{Seuss2016, Roberts2021}. Therefore, as we move to ever deeper and wider spectral line surveys spanning a large spectral bandwidth, 3-dimensional spectroscopic information for some of the most dusty systems in the Universe will become available.

\section*{Acknowledgements}
We would like to thank the anonymous referee for useful comments that improved the manuscript.

The MeerKAT telescope is operated by the South African Radio Astronomy Observatory, which is a facility of the National Research Foundation, an agency of the Department of Science and Innovation. 
We acknowledge the use of the ilifu cloud computing facility – www.ilifu.ac.za, a partnership between the University of Cape Town, the University of the Western Cape, Stellenbosch University, Sol Plaatje University and the Cape Peninsula University of Technology. The ilifu facility is supported by contributions from the Inter-University Institute for Data Intensive Astronomy (IDIA – a partnership between the University of Cape Town, the University of Pretoria and the University of the Western Cape), the Computational Biology division at UCT and the Data Intensive Research Initiative of South Africa (DIRISA).

MJJ acknowledges generous support from the Hintze Family Charitable Foundation through the Oxford Hintze Centre for Astrophysical Surveys. MJJ, IH and AAP acknowledge support of the STFC consolidated grant [ST/S000488/1] and [ST/W000903/1] and MJJ, IH, SMJ and HP from a UKRI Frontiers Research Grant [EP/X026639/1], which was selected by the ERC. IH acknowledges support from the South African Radio Astronomy Observatory which is a facility of the National Research Foundation (NRF), an agency of the Department of Science and Innovation. RPD acknowledges funding by the South African Research Chairs Initiative of the DSI/NRF (Grant ID 77948). AJB acknowledges support from the National Science Foundation through grant AST-2308161 and from the Radcliffe Institute for Advanced Study at Harvard University.
AB, FS and GR acknowledge support from INAF under the Large Grant 2022 funding scheme (project "MeerKAT and LOFAR Team up: a Unique Radio Window on Galaxy/AGN co-Evolution”).
MV acknowledges financial support from the Inter-University Institute for Data Intensive Astronomy (IDIA), a partnership of the University of Cape Town, the University of Pretoria and the University of the Western Cape, and from the South African Department of Science and Innovation's National Research Foundation under the ISARP RADIOSKY2020 and RADIOMAP+ Joint Research Schemes (DSI-NRF Grant Numbers 113121 and 150551) and the SRUG HIPPO Projects (DSI-NRF Grant Numbers 121291 and SRUG22031677).
This research made use of Astropy,\footnote{\url{http://www.astropy.org}} a community-developed core Python package for Astronomy \citep{astropy2013, astropy2018}. This research has made use of CARTA \citep[Cube Analysis and Rendering Tool for Astronomy;][]{comrie2021}. This research has made use of NASA's Astrophysics Data System. This research made use of Montage, which is funded by the National Science Foundation under Grant Number ACI-1440620, and was previously funded by the National Aeronautics and Space Administration's Earth Science Technology Office, Computation Technologies Project, under Cooperative Agreement Number NCC5-626 between NASA and the California Institute of Technology.

\section*{Data Availability}

The MeerKAT visibility data for the MIGHTEE project are available from the SARAO archive under proposal IDs SCI-20180516-KH-01 and SCI-20180516-KH-02. A public release of the reduced image and catalogue products for the first MIGHTEE spectral line data release covering the COSMOS field is imminent (Heywood et al., in prep.).



\bibliographystyle{mnras}
\bibliography{example} 








\bsp	
\label{lastpage}
\end{document}